\documentclass[conference]{IEEEtran}
\usepackage{cite}
\usepackage{amsthm,amsmath,amssymb,amsfonts}
\usepackage[linesnumbered,algoruled,vlined]{algorithm2e}
\usepackage{graphicx}
\usepackage{textcomp}
\usepackage[normalem]{ulem}
\usepackage{xcolor}
\usepackage[labelformat=simple]{subcaption}

\usepackage{booktabs}

\usepackage{hyperref}
\usepackage{cleveref}
\usepackage{flushend}

\newtheorem{remark}{Remark}
\theoremstyle{definition}

\crefname{equation}{Eq.}{Eq.}
\crefname{figure}{Fig.}{Fig.}
\crefname{section}{Sec.}{Sec.}
\crefname{table}{TABLE}{TABLE}
\crefname{algorithm}{Algorithm}{Algorithm}
\crefname{remark}{Remark}{Remark}

\def\BibTeX{{\rm B\kern-.05em{\sc i\kern-.025em b}\kern-.08em
T\kern-.1667em\lower.7ex\hbox{E}\kern-.125emX}}

\SetCommentSty{commfont}
\SetFuncSty{fnfont}

\begin{document}

\title{
  QuIKS: Near-Zero Latency Key Supply with Adaptive Buffering for Resource-Efficient\\ Quantum Key Distribution Networks
}

\author{
  \IEEEauthorblockN{
    Yuxin Chen\IEEEauthorrefmark{1}, Zite Xia\IEEEauthorrefmark{2}, Jian Li\IEEEauthorrefmark{2}\IEEEauthorrefmark{6}, Kaiping Xue\IEEEauthorrefmark{2}\IEEEauthorrefmark{3}\IEEEauthorrefmark{6}, Zhonghui Li\IEEEauthorrefmark{4}, Lutong Chen\IEEEauthorrefmark{2}, Ruidong Li\IEEEauthorrefmark{5}
  }
  \IEEEauthorblockA{\IEEEauthorrefmark{1}Department of EEIS, University of Science and Technology of China, Hefei, Anhui 230027, China}
  \IEEEauthorblockA{\IEEEauthorrefmark{2}School of Cyber Science and Technology, University of Science and Technology of China, Hefei, Anhui 230027, China}
  \IEEEauthorblockA{\IEEEauthorrefmark{3}Hefei National Laboratory, Hefei, Anhui 230088, China}
  \IEEEauthorblockA{\IEEEauthorrefmark{4}School of Electronic and Information Engineering, Anhui University, Hefei, Anhui 230601, China}
  \IEEEauthorblockA{\IEEEauthorrefmark{5}National Institute of Information and Communications Technology, Kanazawa University, Tokyo 184-0015, Japan}
  \IEEEauthorblockA{\IEEEauthorrefmark{6}Corresponding author: J. Li, K. Xue \{lijian9, kpxue\}@ustc.edu.cn}
}

\maketitle

\begin{abstract}
  Quantum key distribution (QKD) networks provide information-theoretically secure keys for distant parties, emerging as a vital alternative to classical cryptography infrastructures threatened by quantum computing.
  In QKD networks, the immediacy of key supply service is crucial to the security and performance of applications, as their data must be encrypted before transmission.
  While key buffering can enable instant key supply services, existing schemes rely on heuristic solutions that incur prohibitive key resource consumption, thus significantly hindering practical deployment.
  To address this issue, we propose QuIKS, an instant key supply scheme based on adaptive buffering, offering the dominant advantage of near-zero key supply latency while consuming ultra-low key resources (i.e., ultra-low buffer size).
  Specifically, it is built upon a novel analytical model that determines the minimum buffer size required to guarantee near-zero-latency key supply performance.
  Guided by this model, QuIKS introduces a lightweight two-phase control algorithm that dynamically determines key relaying requests and adjusts the buffer size by probing real-time application patterns and network conditions.
  Experiments on a real QKD network testbed demonstrate that QuIKS achieves near-zero key supply latency while providing a more than 10-fold reduction in key buffer size compared to state-of-the-art schemes.
\end{abstract}


\section{Introduction}

Quantum key distribution (QKD) \cite{bennett1984quantum} is a cutting-edge technology that provides information-theoretically secure keys for two remote parties based on the principle of quantum mechanics \cite{lo1999unconditional,portmann2022security}, making it a promising solution to defend against the threat posed by quantum computing to classical cryptography \cite{gidney2021factor,shakib2025impersonation}.
To extend these security guarantees from point-to-point links to a large scale, QKD networks have been rapidly developed in recent years \cite{elliott2003quantum,peev2009secoqc,sasaki2011field,chen2021integrated}.
Among various architectural approaches, the trusted-relay technology emerges as the most mature and practical one for building large-scale QKD networks \cite{huttner2022long}.
In this paradigm, a network of relay nodes is interconnected via individual links, in which each pair of adjacent nodes implements a QKD protocol to generate quantum keys.
After that, the end-to-end key can be relayed along a multi-hop path through sequential encryption and decryption between adjacent nodes, consuming shared quantum keys.

The ultimate vision of QKD networks is to serve an on-demand key supply service, seamlessly integrating with applications that currently rely on classical cryptography \cite{cao2022evolution,li2023entanglement,li2025integration}.
However, the QKD network fundamentally struggles to provide this consistent service.
On the one hand, quantum key generation, which determines the key resources between adjacent nodes, is inherently probabilistic and susceptible to environmental disturbances \cite{zhang2025experimental}.
On the other hand, the underlying classical network, essential for key relaying, introduces its dynamics like traffic congestion and network updates \cite{liu2024figret, namjoshi2024algorithms}.
The instability leads to highly volatile network conditions for key relaying, which can cause significant service delays or even failures for secure applications, thereby constituting a major barrier to the widespread adoption of QKD networks.

To address this problem, various efforts have been made to enhance the performance of QKD networks, including sophisticated key management \cite{zhou2022quantum,li2025decentralized}, dynamic routing \cite{mehic2019novel,akhtar2023fast}, and resource allocation \cite{zhang2023routing,zheng2025integration}.
Although these approaches mitigate key generation fluctuations, they still exhibit volatile performance due to heterogeneous paths or strategy switching, and thus, they hardly address the inherent performance fluctuations rooted in the underlying classical networks \cite{zhang2024rond}.
Given the storable nature of keys, a promising approach is to create end-to-end key buffers that can smooth out supply variations from any source.
Existing buffering schemes fall into two categories: those that rely on static pre-allocation mechanisms \cite{cao2019kaas,zhu2023qkd} and those that employ heuristic-based control \cite{stan2025dynamic}.
While these schemes achieve desirable key supply latency, they also consume excessive link-level quantum keys from QKD networks to build up a large buffer size at end nodes.
This results in significant key resource wastage, which cannot be ignored and severely hinders their practical application.

The inability of existing buffer-based schemes to strike an optimal trade-off between key supply performance and resource consumption stems from three fundamental challenges.
\textbf{\textit{(1) The difficulty in modeling the impact of stochastic application requests and QKD networks.}}
The key buffer state is governed by two unpredictable processes: the random arrival of application requests and the fluctuating key relaying process in QKD networks.
This dual source of randomness makes it exceedingly difficult to accurately model and predict buffer state evolution.
\textbf{\textit{(2) The lack of quantitative evaluation for the performance-cost trade-off.}}
Without a rigorous analytical model that connects key supply performance to buffer size, designing a buffer control mechanism becomes a matter of guesswork.
This forces the existing schemes to adopt heuristic strategies with oversized key buffers as a safeguard against exhaustion, which inevitably leads to key resource waste and poor trade-offs.
\textbf{\textit{(3) The low-complexity requirement for the buffer control algorithm.}}
In practice, the key buffer requires a computationally lightweight control algorithm to support real-time management of numerous concurrent key requests.
This requirement creates a dilemma: heuristic strategies maintain computational efficiency but lack accuracy, while sophisticated models achieve high performance but are complex.

To tackle these fundamental challenges, we propose QuIKS, an \uline{\textbf{i}}nstant \uline{\textbf{k}}ey \uline{\textbf{s}}upply scheme based on adaptive buffering with minimal \uline{\textbf{qu}}antum key resource consumption.
The core of QuIKS is an analytical model for the buffer-enabled end-to-end key supply process, aiming to characterize the impact of application requests and QKD networks by unifying both key generation and classical network fluctuations.
Based on this model, we quantify the fundamental trade-off between buffer size and key supply performance using the central limit theorem under the wide-sense stationary assumption, providing theoretical guidance to escape the guesswork of heuristic-based strategies.
Guided by the derived conclusions, we design a lightweight buffer control algorithm that dynamically adapts to application patterns and network conditions in a real-time manner, relaxing the stationary assumption.
We then implement and evaluate QuIKS on a real-world QKD network testbed.
Experimental results demonstrate that QuIKS achieves a near-zero key supply latency, while providing a more than 10-fold reduction in key buffer size compared to state-of-the-art schemes.
By these efforts, QuIKS paves the way for the widespread adoption of QKD networks.

The contributions of this work are summarized as follows:
\begin{itemize}
  \item For the first time, we propose a novel theoretical model for key supply service, which can quantify the mathematical relationship between buffer size and key supply performance, providing theoretical guidance for optimal buffer control design.
  \item We design a lightweight, theory-guided adaptive buffer control algorithm, named QuIKS, that translates our analytical insights into a practical mechanism for providing instant key supply service with ultra-low key resource consumption.
  \item We conduct extensive experiments on a real QKD network testbed, demonstrating that QuIKS reduces key buffer size by over 90\% compared to state-of-the-art schemes while achieving near-zero key supply latency.
\end{itemize}

The rest of this paper is organized as follows.
First, we present the background and our motivation in \cref{bg_motiv}.
In \cref{model}, we detail the system model and the analytic result.
After that, we describe the QuIKS design in \cref{algo}.
Then, in \cref{expresult}, we introduce the implementation and conduct experiments on a real testbed.
Finally, we briefly review related work in \cref{works} and conclude our work in \cref{conclu}.

\section{Background and Motivation}\label{bg_motiv}

In this section, we first introduce the necessary preliminaries and then articulate our motivation with an experiment.
After that, we present our design goals.

\subsection{Key Supply Services in QKD Networks}

QKD networks \cite{elliott2003quantum,peev2009secoqc,sasaki2011field,chen2021integrated} are built to extend the range and scalability of various QKD protocols \cite{du2024twin,zhuang2025ultrabright}, which are inherently limited by quantum signal attenuation in the physical channels.
Functionally, the architecture of QKD networks is standardized into a three-layer framework \cite{y3802}.
The bottom layer comprises multiple independent point-to-point QKD protocols responsible for generating quantum keys between respective node pairs.
The middle layer handles end-to-end key supply services for secure applications (e.g., IPSec \cite{gao2025ipseq} or TLS \cite{garcia2025enhanced}) in the top layer by leveraging these quantum keys to perform key relay between two distant nodes, denoted as a source node $s$ and a destination node $d$.
A multi-hop path connecting $s$ and $d$ is first determined, and the end-to-end key originates at $s$.
Between each adjacent node pair, the end-to-end key is encrypted with a quantum key generated by the bottom layer at one node and then transmitted to the next node via a classical channel, where the same quantum key is used for decryption.
The one-time pad (OTP) algorithm is used to ensure information-theoretic security \cite{shannon1949communication}.
The process is repeated along the path until the end-to-end key reaches $d$.
Finally, $d$ acknowledges the success of the relay, and $s$ supplies the end-to-end key to the requesting application.

\subsection{Motivation}
Due to the complicated process required to supply end-to-end keys, QKD networks struggle to provide an instant key supply service.
At the bottom layer where QKD protocols are performed, secure key generation is governed by the probabilistic nature of quantum mechanics and the post-processing performance over the classical channel \cite{mehic2017analysis} and thus exhibits significant fluctuations, as amply demonstrated in operational QKD networks \cite[Fig. 5]{martin2024madqci}.
The relaying process, which relays end-to-end keys hop-by-hop over a classical network, is exposed to the full spectrum of classical network dynamics in either the current overlay or the future integrated architecture \cite{li2025integration}.
Consequently, the key supply service relies on a cascade of dynamic processes, leading to highly fluctuating latency for key requests from applications.

\begin{figure}
  \centering
  \begin{minipage}[t]{0.48\linewidth}
    \centering
    \includegraphics[width=\linewidth]{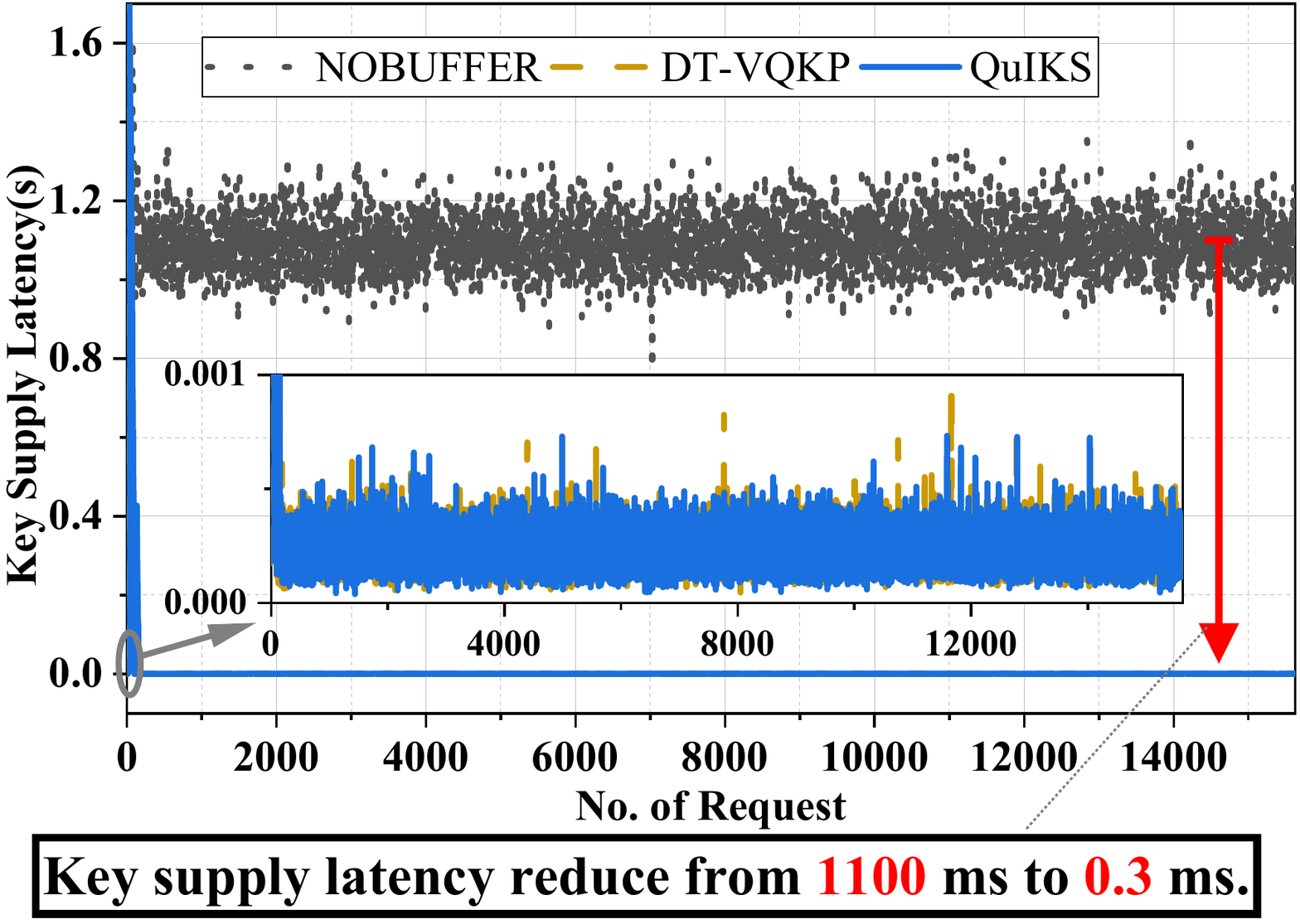}
    \subcaption{Key supply latency.}\label{fig_motivdelay}
  \end{minipage}~
  \begin{minipage}[t]{0.48\linewidth}
    \centering
    \includegraphics[width=\linewidth]{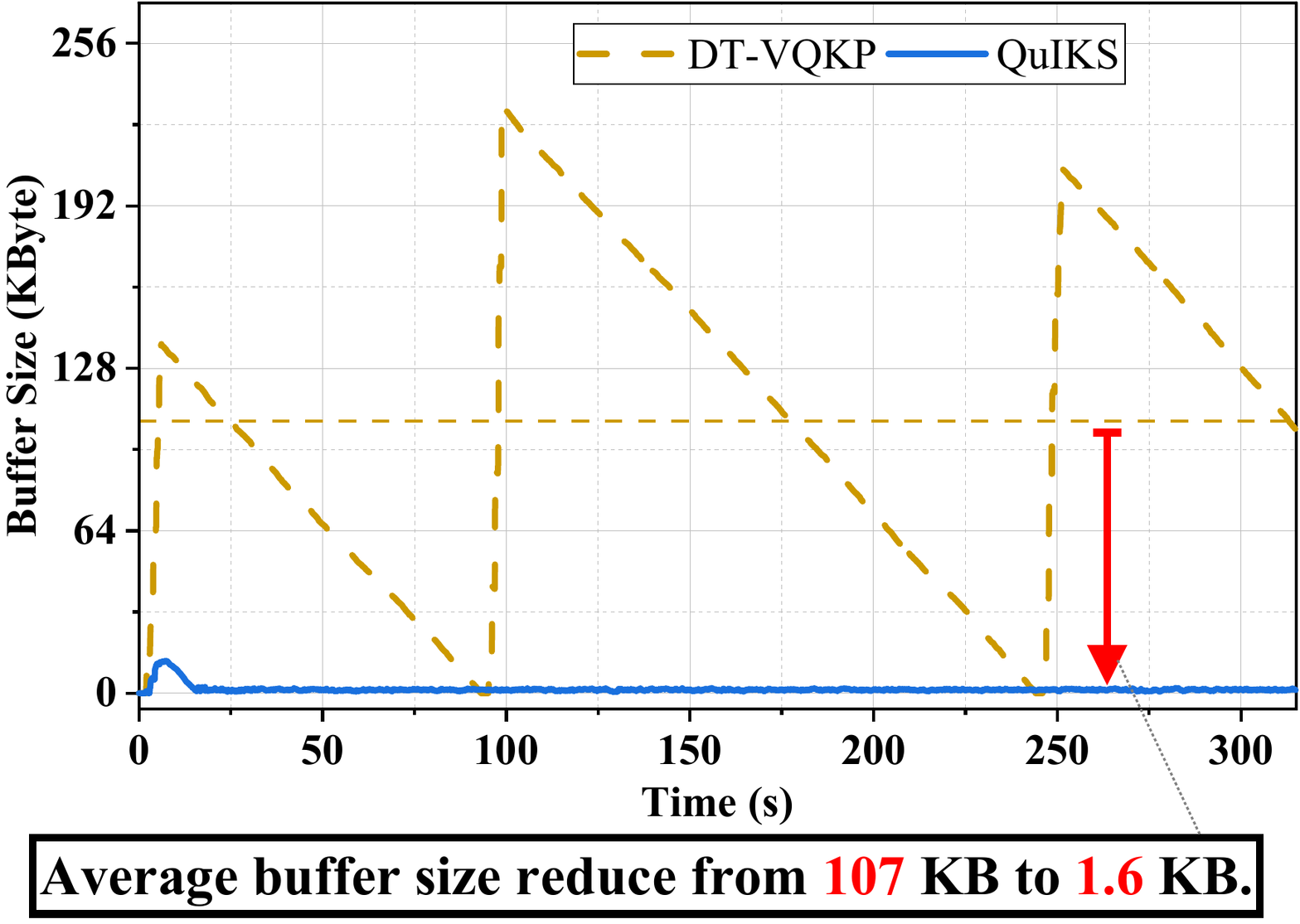}
    \subcaption{Buffer size.}\label{fig_motivbuf}
  \end{minipage}
  \caption{Performance of the buffer-enabled key supply schemes.}\label{fig_pre}
\end{figure}

While many existing works \cite{zhou2022quantum,li2025decentralized,mehic2019novel,akhtar2023fast,zhang2023routing,zheng2025integration} focus on mitigating quantum layer performance fluctuations through key management or multipath key aggregation, they also introduce new fluctuations due to heterogeneous paths or strategy switching.
Coupled with the inherent dynamics of the underlying classical network, the key relaying process and the key supply performance remain fluctuating.
\cref{fig_motivdelay} depicts the test results obtained from a 14-node QKD network testbed (detailed in \cref{subsec_impl}), revealing extremely high and unstable key supply latency.
The unstable latency leads to lags or interruptions in the key supply, resulting in degraded data transmission performance or compromised security, hindering the widespread application of QKD networks.

Some pioneering studies \cite{cao2019kaas,stan2025dynamic} already consider the storable nature of keys and introduce end-to-end key buffering by relaying keys in advance, thereby providing a near-zero latency key supply service.
As shown in \cref{fig_motivdelay}, the state-of-the-art solution, DT-VQKP \cite{stan2025dynamic}, as well as QuIKS (formally proposed in \cref{algo}), demonstrate a decisive advantage of 0.3 ms versus 1100 ms in key supply latency and low fluctuation over direct relaying without a buffer.
Despite this attractive advantage, DT-VQKP also requires a large buffer size of 107 KByte on average, as depicted in \cref{fig_motivbuf}, due to heuristic thresholds and a bursty buffer filling strategy.
Each buffered key must consume the same number of quantum keys on each link along the relaying path in QKD networks and cannot be reused for any other node pairs but the dedicated one.
This fact leads to severe wastage of key resources, thereby degrading network performance and significantly hindering its value in practical QKD networks, which are key-limited with a typical quantum key generation rate of 79.3 Kbps \cite{chen2021integrated}.
However, it is not always necessary to have such a huge size. As also depicted in \cref{fig_motivbuf}, by maintaining a buffer size of 1.6 KByte, the proposed QuIKS still achieves optimal key supply performance.

\subsection{Design Goals}
The above results reveal that a buffer can effectively mitigate fluctuations in key supply performance, but a large buffer poses a significant resource burden.
Therefore, it is necessary to design a resource-efficient scheme while achieving the same key supply performance as in practical QKD networks.
However, designing such a scheme remains challenging, primarily due to the stochastic nature of application requests and QKD networks, which simultaneously influence the buffer dynamics, and the lack of a quantitative evaluation for the performance-key consumption trade-off. These challenges, along with the lightweight requirement for practical algorithms, are the main reasons existing schemes employ heuristic control.
By addressing these challenges, a practical buffering-based scheme should meet the following design goals:
\begin{enumerate}
 \item \textbf{Guaranteed Performance}: To achieve continuous instant key supply for applications.
 \item \textbf{Minimal Buffer Size}: To operate with the minimal possible key resource consumption.
 \item \textbf{Lightweight and Adaptive Control}: To enable real-time, adaptive control with low computational overhead.
\end{enumerate}

\section{Buffer-enabled Key Supply Model}\label{model}

In this section, we first model the key supply with buffering as a special queuing system in QKD networks. Then, we analyze this system and derive theoretical conclusions to guide the key supply scheme proposed in the next section.

\subsection{System Model}

\begin{figure}[tpb]
  \centering
  \includegraphics[width=\linewidth]{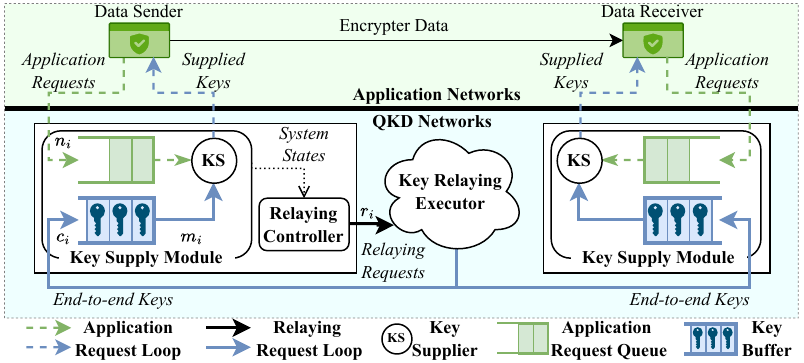}
  \caption{Buffer-enabled key supply model in QKD networks.}\label{fig_bufmodel}
\end{figure}

In a QKD network, considering a single data transmission direction for a pair of secure applications, key requests are always actively initiated by the sender, while the receiver passively generates key requests based on the received encrypted data.
Therefore, the dynamics of the pair of key buffers are dominated by the sender-side one.
Based on the basic key supply service in QKD networks, a buffer-enabled end-to-end key supply model is shown in \cref{fig_bufmodel}.
On both sides, application requests are decoupled from the key relaying process by a buffer-enabled key supply module, which is a parallel queuing system constituted by an application request queue and a key buffer.
In this system, a key supplier operates according to the following rule: responding to a queuing application request by retrieving a key from the key buffer. Since this operation is simple, the service time is assumed to be zero.
Additionally, to fill the key buffer, a relaying controller sends relaying requests according to the system states on the sender side.

Aligning with the existing QKD networks \cite{dervisevic2025quantum}, the model assumes that key requests and relaying operate in fixed-size blocks.
According to cutting-edge protocol design \cite{chen2025asynchronous}, the relaying process can operate reliably with no resource wastage. Thus, it assumes every relaying request can receive an end-to-end key.
In this context, the relaying delay, which is the time between sending a relaying request and receiving the corresponding end-to-end key, can uniformly describe all QKD network fluctuations caused by key management schemes or classical networks.
To precisely describe the changes in the key buffer, we introduce a time slot model, where each time slot has a length of $T$.
On this basis, \cref{tab_notation} presents the definitions of the used notations, and the state transition equations for the key buffer are as follows:
\begin{align}
  &m_{i}=m_{i-1}-n_{i}+c_{i}, \label{eq_buftrans}\\
  &c_i=\sum_{j=1}^{K}{r_{i-j}\omega_{i-j,j}},\label{eq_bufstate}
\end{align}
where $\omega_{i,j}$ is the probability that a relaying request sent in the $i$-th time slot experience a $j$-time slot delay, thus satisfying $\sum_{j=1}^{K}{\omega_{i,j}}=1$.
\cref{eq_bufstate} reveals how QKD network fluctuations affect the arrival of keys.
According to the service rule, at least one of the queues must be empty at the end of a time slot, as the service time is zero.
Therefore, $m_i$ physically indicates the key buffer size when it is positive, while its absolute value indicates a backlog of application requests vice versa.

\begin{table}[t]
  \centering
  \caption{Notations}\label{tab_notation}
  \setlength{\aboverulesep}{1pt}
  \setlength{\belowrulesep}{1pt}
  \begin{tabular}{lp{6.9cm}}
    \toprule
    Notation & Definition\\
    \midrule
    $m_i$            & The key buffer size at the end of the $i$-th time slot.\\
    $n_i$            & The number of application requests arriving during the $i$-th time slot.\\
    $c_i$            & The number of keys arriving at the key buffer during the $i$-th time slot.\\
    $r_i$            & The number of relaying requests sent by the controller during $i$-th time slot. \\
    $\omega_j$       & The probability that a relaying request receives the corresponding key after exactly $j$ time slots.\\
    $K$              & The maximum possible delay of relaying requests in time slots, such that $\omega_j = 0$ for all $j > K$.\\
    % $W_k$            & The queuing time of $k$-th application request before it is satisfied.\\
    % $t^n_k,t^c_k$    & The arrival time point of $k$-th application request ($t^n_k$) and $k$-th relayed key ($t^c_k$), respectively.\\
    $\mu_n,\mu_c$    & The long-term mean number of arrived application requests ($n_i$) and keys ($c_i$) per time slot, respectively. (i.e., $\mu_n = \lim_{N\rightarrow\infty}\frac{1}{N}\sum_{i=1}^{N}{n_i}$ and $\mu_c = \lim_{N\rightarrow\infty}\frac{1}{N}\sum_{i=1}^{N}{c_i}$).\\
    $\Delta_{x,i}$   & The change in the key buffer size $m$ from the end of slot $x$ to the end of slot $i$, i.e., $m_i - m_x$\\
    $C_n(x)$         & The auto-covariance function of the series $\{n_i\}$ at lag $x$.\\
    \bottomrule
  \end{tabular}
\end{table}

\subsection{Analysis of Relaying Control Strategy}\label{subsec_strategy}

To optimize key supply performance, it is equivalent to minimizing the queuing time of application requests before they are satisfied.
When it is almost surely that the key buffer is non-empty when each application request arrives, the queuing time of application requests is 0.
In the proposed model, it should satisfy that $m_{i} > 0$ for all $i$.
According to \cref{eq_buftrans}, $m_i$ effectively constitutes a random walk.
Assuming that the long-term mean number of arrived application requests and keys per time slot exist, the asymptotic behavior of $m_i$ satisfies the following equation:
\begin{align}
  \lim_{N\rightarrow\infty}\frac{m_N}{N}&=\lim_{N\rightarrow\infty}\frac{1}{N}{\sum_{i=1}^{N}{c_i}}-\lim_{N\rightarrow\infty}\frac{1}{N}{\sum_{i=1}^{N}{n_i}}=\mu_c-\mu_n.\nonumber
\end{align}
When $\mu_c<\mu_n$, which means that keys arrive slower than application requests, $\lim_{N\rightarrow\infty}{m_N}=-\infty$.
In this case, since $n_i$ is non-negative, $m_{i-1}\ge n_{i}$ is unachievable.
When $\mu_c>\mu_n$, although there is $\lim_{N\rightarrow\infty}{m_N}=+\infty$ and thus $m_{i-1}\ge n_{i}$ definitely holds, the key buffer size will be unbounded.
This result implies that the system cannot reach a steady state, rendering control meaningless, and an unbounded buffer contradicts our design goals.

This analysis reveals that for the system to operate in a stable regime with a bounded buffer, the primary principle for determining $r_i$ must satisfy the condition $\mu_c = \mu_n$.
Assume the relaying delay is stationary, i.e., $\omega_{i,j} = \omega_j$, meaning that network conditions do not exhibit significant changes over the considered timescale.
If application requests depend only on their respective workloads, $n_i$ and $\omega_{j}$ are independent.
Given a reactive relaying control strategy of $r_i=n_i$, $\mu_c$ can be calculated according to \cref{eq_bufstate} as follows:
\begin{align}
  \mu_c&=\lim_{N\rightarrow\infty}\frac{1}{N}\sum_{i=1}^{N}{\sum_{j=1}^{K}{n_{i-j}\omega_{i-j,j}}}\nonumber\\
  &=\sum_{j=1}^{K}{\omega_{j}\lim_{N\rightarrow\infty}\frac{1}{N}\sum_{i=1}^{N}{n_{i-j}}}=\sum_{j=1}^{K}{\omega_{j}}\mu_n\nonumber.
\end{align}
Recalling that $\sum_{j=1}^{K}{\omega_{j}}=1$, there holds $\mu_c=\mu_n$.

\begin{remark}\label{remark_relay}
  \uline{The reactive strategy, $r_i = n_i$, ensures long-term rate matching, which is the necessary condition to prevent both persistent key depletion and an unbounded buffer size.}
\end{remark}

\subsection{Analysis of the Optimal Key Buffer Size}\label{subsec_buffer}

The reactive strategy, $r_i=n_i$, prevents the first-order systematic drift in the key buffer.
However, to achieve instant key supply, i.e., $m_i > 0 $, it is necessary to account for short-term fluctuations, which requires a higher-order analysis of buffer size dynamics.
By expanding \cref{eq_buftrans}, $m_i$ can be expressed as follows:
\begin{align}
  m_i&=m_x-n_{x+1}+c_{x+1}-\cdots-n_i+c_i \nonumber\\
  &=m_x-\sum_{l=x+1}^{i}{n_l}+\sum_{l=x+1}^{i}{\sum_{j=1}^{K}{r_{l-j}\omega_{l-j,j}}}.\label{eq_kbs1}
\end{align}
Applying the reactive strategy $r_i=n_i$, substituting the stationary condition of $\omega_{l-j,j}$, and rearranging \cref{eq_kbs1} yields:
\begin{align}
  \Delta_{x,i}&=-\sum_{j=1}^{K}{\omega_{j}\sum_{l=x+1}^{i}{n_l}}+\sum_{j=1}^{K}{\omega_{j}\sum_{l=x+1}^{i}{n_{l-j}}} \nonumber\\
  &=\sum_{j=1}^{K}{\omega_{j}{\sum_{l=x+1}^{i}{n_{l-j}-n_l}}}=\sum_{j=1}^{K}{\omega_{j}S_{x,i}(j)},\label{eq_kbs_final}
\end{align}
where $S_{x,i}(j)=\sum_{l=x+1}^{i}{n_{l-j}-n_l}$.

\cref{eq_kbs_final} implies that under the reactive strategy, the change in the key buffer size is only related to the delay characteristics $\omega_j$ and $K$ of QKD networks, and the application request arrival characteristics $n_i$.
Further assuming the application request arrival is a wide-sense stationary (WSS) process over the considered timescale, which indicates that $E[n_i]$ exists and is constant, it is clear that $E[S_{x,i}(j)]=0$, and thus $E[\Delta_{x,i}]=0$.
This result means that the key buffer size will fluctuate around a fixed value, and the distribution of $\Delta_{x,i}$ determines the key buffer size required to ensure $m_{i-1}\ge n_i$.

Examining the variance of $\Delta_{x,i}$, i.e., $\sigma_{\Delta}$, it is effectively the second moment of $\Delta_{x,i}$, expressed as follows:
\begin{align}
  \sigma_{\Delta}^2 &= E[ ( \sum_{j=1}^{K} \omega_j S_{x,i}(j) )^2 ]=\sum_{j=1}^{K} \sum_{k=1}^{K} \omega_j \omega_k E[S_{x,i}(j) S_{x,i}(k)].\nonumber
  % \label{eq_varm}
\end{align}
Let $S_{x,i}(j)=Z_{x}(j)-Z_{i}(j)$, where $Z_{x}(j)=\sum_{p=x-j+1}^{x}{n_p}$, and substituting it into $E[S_{x,i}(j) S_{x,i}(k)]$ yields:
\begin{equation}
  \begin{aligned}
    E[S_{x,i}(j)S_{x,i}(k)]&= E[Z_x(j)Z_x(k)]-E[Z_x(j)Z_i(k)]\\
    &-E[Z_i(j)Z_x(k)]+E[Z_i(j)Z_i(k)].
  \end{aligned}\label{eq_fourterm}
\end{equation}
Since $n_i$ is WSS, its auto-covariance depends only on the time difference and $\mu_n=E[n_i]$. Hence, $E[n_pn_q]=\mu_n^2+C_n(p-q)$, and $E[Z_x(j)Z_i(k)]$ has the following form:
\begin{align}
  E[Z_x(j)Z_i(k)]&= jk\mu_n^2+\sum_{p=x-j+1}^{x}\sum_{q=i-k+1}^{i}C_n(p-q). \nonumber
\end{align}
Further, assume that $i-x$ is sufficiently large compared to the typical correlation timescale of the application request. This is a reasonable assumption for many request models, which often exhibit short-range dependence or an exponentially decaying auto-correlation function, leading to $C_n(p-q) \approx 0$ for $p$ near $x$ and $q$ near $i$.
On this basis, summing all the four terms of \cref{eq_fourterm} and simplifying yields:
\begin{align}
  E[S_{x,i}(j)S_{x,i}(k)] = 2 \sum_{p=1}^{j} \sum_{q=1}^{k} C_n(p-q).\nonumber
  % \label{eq_ess}
\end{align}
Let $\Lambda(j,k)=\sum_{p=1}^{j} \sum_{q=1}^{k} C_n(p-q)$ for simplification, $\sigma_{\Delta}^2$ can be finally determined as follows:
\begin{align}
  \sigma_{\Delta}^2 &= 2\sum_{j=1}^{K}{\sum_{k=1}^{K}\omega_j \omega_k \Lambda(j,k)}.\label{eq_varfinal}
\end{align}

While the exact distribution of $\Delta_{x,i}$ is unknown, its nature as a sum of many random variables justifies an approximation using the central limit theorem.
Therefore, $\Delta_{x,i}$ is assumed to follow the normal distribution $\mathcal{N}(0, \sigma_{\Delta}^2)$.
The objective of $m_i>0$ is to dimension an initial buffer size $L$ such that $L+\Delta_{x,i} >0$ with a small tolerance $\epsilon$.
It leads to the condition $P(\Delta_{x,i} < -L) \le \epsilon$, and thus $L$ can be given by the solution to $\Phi(-L/\sigma_{\Delta}) = \epsilon$, where $\Phi$ is the standard normal CDF.

\begin{remark}
  \uline{For a WSS application request arrival process under the reactive strategy, the initial buffer size $L$ satisfying $\Phi(-L/\sigma_{\Delta})=\epsilon$ provides an approximate guarantee against lagged key supply with a probability of at least $1-\epsilon$.}
  \label{remark_buffer}
\end{remark}

\section{Lightweight Buffer Control Algorithm}\label{algo}

\subsection{Overview}
Based on the preceding model and theoretical results, this section proposes a lightweight control algorithm, named QuIKS, for practice.
The critical challenge is that the target buffer size needs to be calculated with unknown and dynamic system parameters, and QuIKS addresses this in a two-phase manner.
It begins with a \textbf{\textit{Probing and Adjusting}} phase to online estimate system parameters and establish a buffer guided by \cref{remark_buffer}.
Subsequently, it enters a \textbf{\textit{Stable Controlling}} phase, efficiently applying the reactive strategy as \cref{remark_relay} describes.
To maintain adaptability to the non-stationary conditions in QKD networks, a lightweight mechanism reinitiates probing only when a significant buffer size drop is detected, ensuring robust and instant key supply.

\subsection{Probing and Adjusting Phase}

\begin{algorithm}[tbp]
  \small
  \caption{Probing and Adjusting Phase}\label{alg_probe}
  \KwIn{The application requests record list $\{n\}$;
    The relaying delay record list $\{w\}$;
    Current time slot $i_c$.
  }
  Initiate $i_s\leftarrow i_c$, $\{n\}\leftarrow\{0\}$, $\{w\}\leftarrow\{0\}$, $K\leftarrow\infty$\;
  \While(\tcp*[f]{Probe parameters}){$i_c<(i_s+(\alpha+1) K)$}{\label{alg_line_probing}
    $N \leftarrow$ \FuncSty{Num\_Of\_Rcved\_App\_Requests()}\;
    $n_{i_c-i_s}\leftarrow N$ for list $\{n\}$\;
    Send $N$ relaying requests\;
    \If{$i_c<(i_s+\alpha K)$}{
      Send additional $\beta N$ relaying requests\;
    }
    \ForAll{$w_i$ in $\{w\}$}{
      $W \leftarrow$ \FuncSty{Num\_Of\_Rcved\_Keys\_With\_Delay($i$)}\;
      $w_{i}\leftarrow w_{i}+W$\;
    }
    \If{All requests sent in any time slot are satisfied}{
      $K\leftarrow$ Index of the largest non-zero element of $\{w\}$\;
    }
    Enter next time slot and $i_c=i_c+1$\;
  }
  Calculate $\sigma_{\Delta}^2$ based on $\{n\}$, $\{w\}$, and $K$ with \cref{eq_varfinal}\;
  $M\leftarrow$\FuncSty{Size\_Of\_Key\_Buffer()}\;
  $d\leftarrow [5\sigma_{\Delta}]-M$\;
  \While(\tcp*[f]{Adjust buffer size}){$d\ne0$}{\label{alg_line_adjust}
    $N \leftarrow$ \FuncSty{Num\_Of\_Rcved\_App\_Requests()}\;
    Send $\max(0,N+d)$ relaying requests\;
    $d\leftarrow d+N-\max(0,N+d)$\;
    Enter next time slot and $i_c=i_c+1$\;
  }
  Enter \cref{alg_stable} steady controlling phase\;
\end{algorithm}

\cref{alg_probe} illustrates the operation of the \textit{\textbf{Probing and Adjusting}} phase of QuIKS.
This phase comprises two processes: probing parameters (line \ref{alg_line_probing}) and adjusting buffer size (line \ref{alg_line_adjust}).

The probing process lasts for $(\alpha+1) K$ time slots, during which relaying requests are sent at a rate of $\beta+1$ times the arrival rate of application requests in the first $\alpha K$ time slots.
A larger $\alpha$ extends the observation window, perhaps yielding a more accurate estimation of $C_n(x)$.
A larger $\beta$ may result in a more reliable measurement of the maximum delay $K$ and the delay distribution $\omega_j$.
In this process, the algorithm continuously records the number of received application requests in each time slot and the number of received keys corresponding to each delay value.
Based on the records, after the relaying requests sent in a certain time slot are fully satisfied, $K$ is updated by finding the maximum recorded relaying delay, which ensures that $K$ is continuously revised monotonically. This process will definitely end as long as $K$ is finite.
Subsequently, $\{w\}$ is normalized to obtain each $\omega_j$, and then $\sigma_{\Delta}$ can be calculated according to \cref{eq_varfinal}.

The adjusting process aims to steer the mean buffer level to a target size, which is set at $5\sigma_{\Delta}$ to ensure a small lagged supply probability ($\epsilon < 10^{-6}$) according to \cref{remark_buffer}.
It first calculates the required adjustment $d$, which is the difference between the target size and the current buffer size $M$.
QuIKS uses the instantaneous buffer size $M$ as an estimation of the pre-adjustment mean, which is justified because the last $K$ time slots of the probing process intentionally wait for the over-requested keys to arrive, making the buffer enter a steady state.
The algorithm then closes the difference $d$ by either over-sending ($d>0$) or under-sending ($d<0$) relaying requests in several subsequent time slots.
Ultimately, the buffer size is centered around the new target mean of $5\sigma_{\Delta}$ before entering the stable controlling phase.

\subsection{Stable Controlling Phase}

\begin{algorithm}[tbp]
  \small
  \caption{Stable Controlling Phase}\label{alg_stable}
  \KwIn{The calculated $\sigma_{\Delta}^2$;
    Current time slot $i_c$.
  }
  \While{True}{
    $M\leftarrow$\FuncSty{Size\_Of\_Key\_Buffer()}\;
    \If{$M<\sigma_{\Delta}$}{
      Enter \cref{alg_probe} probing and adjusting phase\;
      \Return\;
    }
    $N \leftarrow$\FuncSty{Num\_Of\_Rcved\_App\_Requests()}\;
    Send $N$ relaying requests\;\label{alg_line_stablesend}
    Enter next time slot and $i_c=i_c+1$\;
  }
\end{algorithm}

\cref{alg_stable} presents the operation of the \textit{\textbf{Stable Controlling}} phase of QuIKS.
In this phase, the task is to simply execute the reactive control strategy described as \cref{remark_relay}, which always sends the same number of relaying requests as that of received application requests (line \ref{alg_line_stablesend}).
While this simple strategy is highly efficient, QuIKS should remain adaptive to non-stationary network conditions without incurring significant overhead.
To achieve this, it employs a lightweight, trigger-based mechanism that initiates a new probing phase only if the buffer size $M$ drops below a conservative threshold of $\sigma_{\Delta}$.
Given that the buffer is maintained at a target of $5\sigma_{\Delta}$ and approximately follows $\mathcal{N}(5\sigma_{\Delta},\sigma_{\Delta})$, such a significant drop is a low-probability ($< 10^{-4}$) event under stable conditions, making it an efficient indicator of significant shifts in application requests or network characteristics.
This ensures that while minor, harmless fluctuations are ignored, the system remains robust by maintaining a sufficient buffer ($M \ge \sigma_{\Delta}$) to provide a continuous instant key supply service.

\subsection{Complexity Analysis}
\subsubsection{Space Complexity}
\cref{alg_probe} collects the number of application requests and relay delays for $(\alpha+1)K$ time slots, while the peak buffer size is $(\alpha\beta -1)K\mu_n$.
Thus, its space complexity is $O(K\mu_n)$. The space complexity of \cref{alg_stable} is $O(1)$ owing to its simple design.

\subsubsection{Time Complexity}
For \cref{alg_probe}, the complexity per time slot is $O(1)$, as the operations within each slot are limited to constant-time tasks such as comparisons and data recording.
The primary computational overhead from calculating $\sigma_{\Delta}^2$, which has an overall complexity of $O(K^2)$.
This process involves computing $C_n(x)$ in $O(K^2)$ time and then constructing the $\Lambda(j,k)$ matrix.
By employing the dynamic programming recurrence $\Lambda(j,k) = \Lambda(j-1, k) + \Lambda(j, k-1) - \Lambda(j-1, k-1) + C_n(j-k)$, the matrix construction and the subsequent double summation for $\sigma_{\Delta}^2$ are both efficiently performed in $O(K^2)$.
Thus, the overall time complexity of \cref{alg_probe} is $O(K^2)$, and the time complexity of \cref{alg_stable} is $O(1)$.

\begin{figure}[tpb]
  \begin{minipage}[t]{\linewidth}
    \centering
    \includegraphics[width=0.88\linewidth]{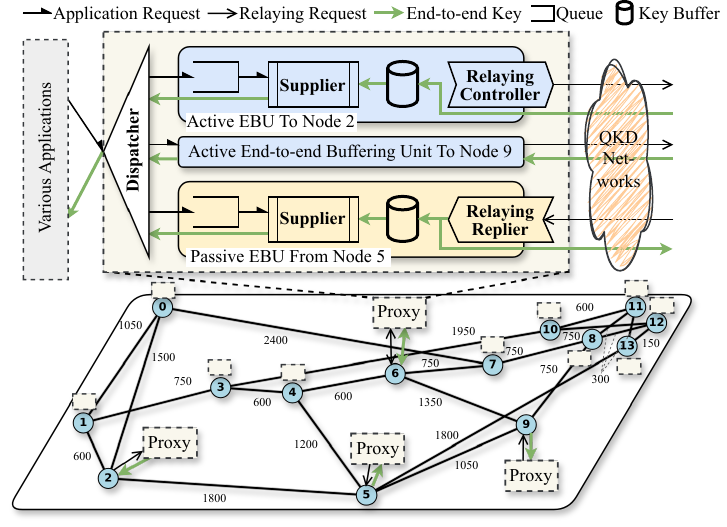}
    \caption{Implementation of the testbed and key buffering architecture.}\label{fig_impl}\label{fig_topo}
  \end{minipage}
\end{figure}

\section{Experiment Results}\label{expresult}

In this section, we introduce the implementation of the QKD network testbed and key buffering architecture. After that, we conduct the experiments to evaluate QuIKS.

\subsection{Implementation}\label{subsec_impl}

To evaluate QuIKS in a more realistic environment, we construct a QKD network testbed.
We utilize 14 small-scale routers to form an NSFnet topology, with adjacent routers being directly connected via Ethernet cables, as shown in \cref{fig_impl}, where the numbers on the links represent the routing metrics.
A forwarder daemon is deployed on each router, implementing the cutting-edge asynchronous key relay protocol \cite{chen2025asynchronous}.
Forwarders on adjacent routers establish TCP connections to form an overlay QKD network.
Building upon this, we further design a proxy daemon to achieve application-transparent key buffering.
This proxy acts as a middleware, communicating with applications and the forwarder via inter-process communication methods.
Each proxy incorporates a request dispatcher and several end-to-end buffering units (EBU).
The dispatcher redirects requests to the correct EBU based on the types-address tuple of encryption-destination or decryption-source, and creates an EBU when it does not exist.
The EBU is responsible for request queuing, key supplying, and managing the key buffering with schemes like QuIKS.

\subsection{Experiment Settings}

In the experiments, we replay the data collected from commercial QKD devices (model: QuantumCTek QKD-PHA1250-S) to simulate realistic key generation between adjacent nodes.
We use a 256-bit key block size to align with the key length of the AES algorithm, following \cite{chen2025asynchronous,stan2025dynamic}.
The comparison schemes include KaaS \cite{cao2019kaas} with fixed relaying rates of 40 requests per second (rps) and 120 rps, as well as the state-of-the-art ST-VQKP and DT-VQKP, with their default parameters as specified in \cite{stan2025dynamic}.
For QuIKS, we set $\alpha = \beta = 2$ as an empirical compromise.
The duration of a time slot is set to 50 ms.
Experimental variables include the application request rate, distribution, and total key demand, as well as the available quantum keys and link delay in the QKD network.
Link delay is simulated using the \texttt{tc-netem} tool, following a normal distribution $\mathcal{N}(x, \frac{x}{10})$, where $x$ is the mean link delay.

The considered performance metrics include buffer size, key supply latency, instant key supply ratio, and application completion ratio.
The key supply latency refers to the time between an application sending a request and receiving the corresponding key.
Note that the latency involves inherent processing time on the real system, but its optimization is beyond the scope of this paper.
The instant key supply ratio is the proportion of requests with a key supply latency $<1$ ms among all requests, where the threshold is to exclude the inherent processing time.
The application completion ratio is the proportion of applications for which all requests are satisfied, relative to the total number of applications.

\subsection{Performance in a Single-Application Scenario}

\begin{figure}[t]
  \centering
  \begin{minipage}[t]{0.48\linewidth}
    \centering
    \includegraphics[width=\linewidth]{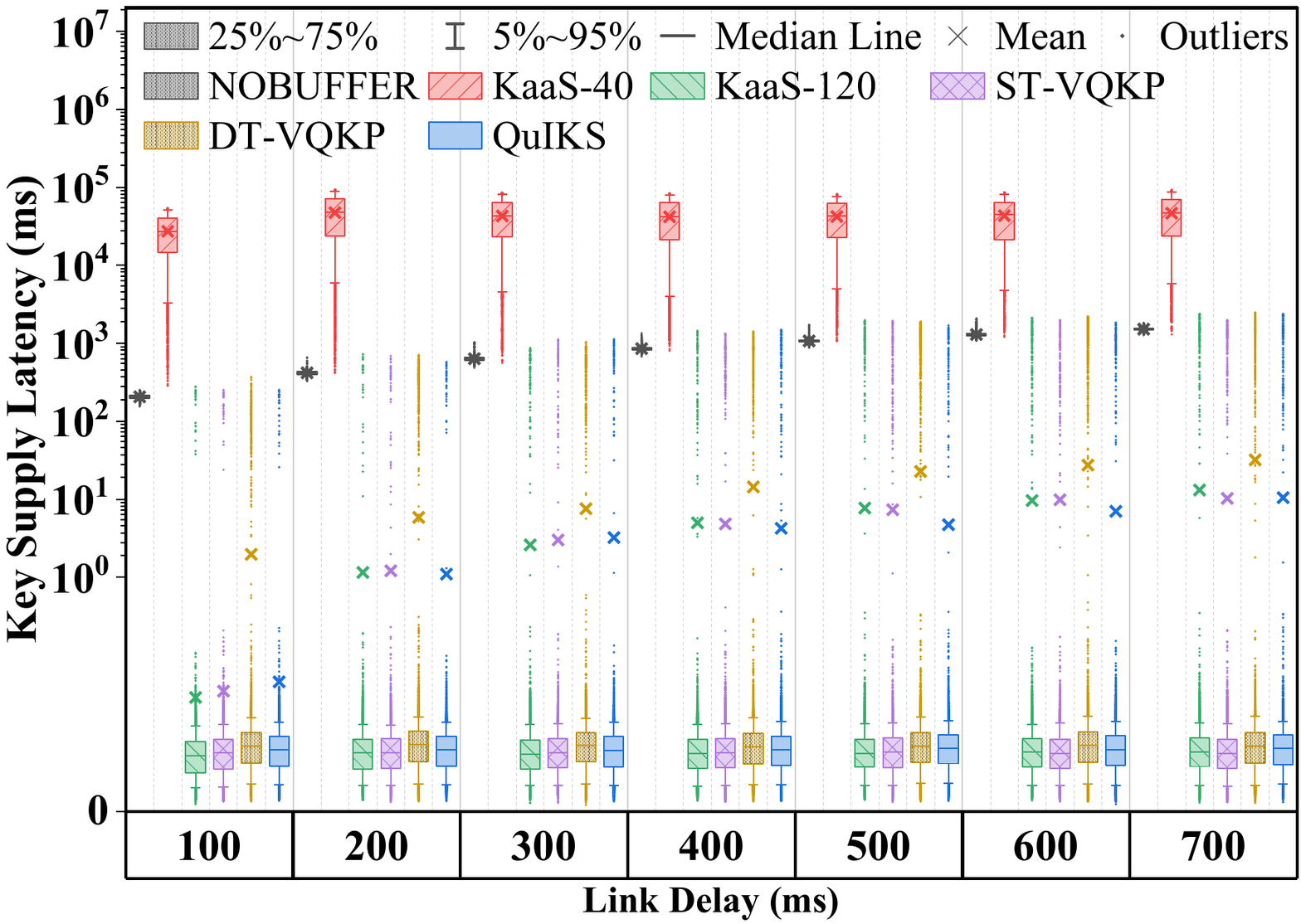}
    \subcaption{Key supply latency.}\label{fig_ksdpoi}
  \end{minipage}~
  \begin{minipage}[t]{0.48\linewidth}
    \centering
    \includegraphics[width=\linewidth]{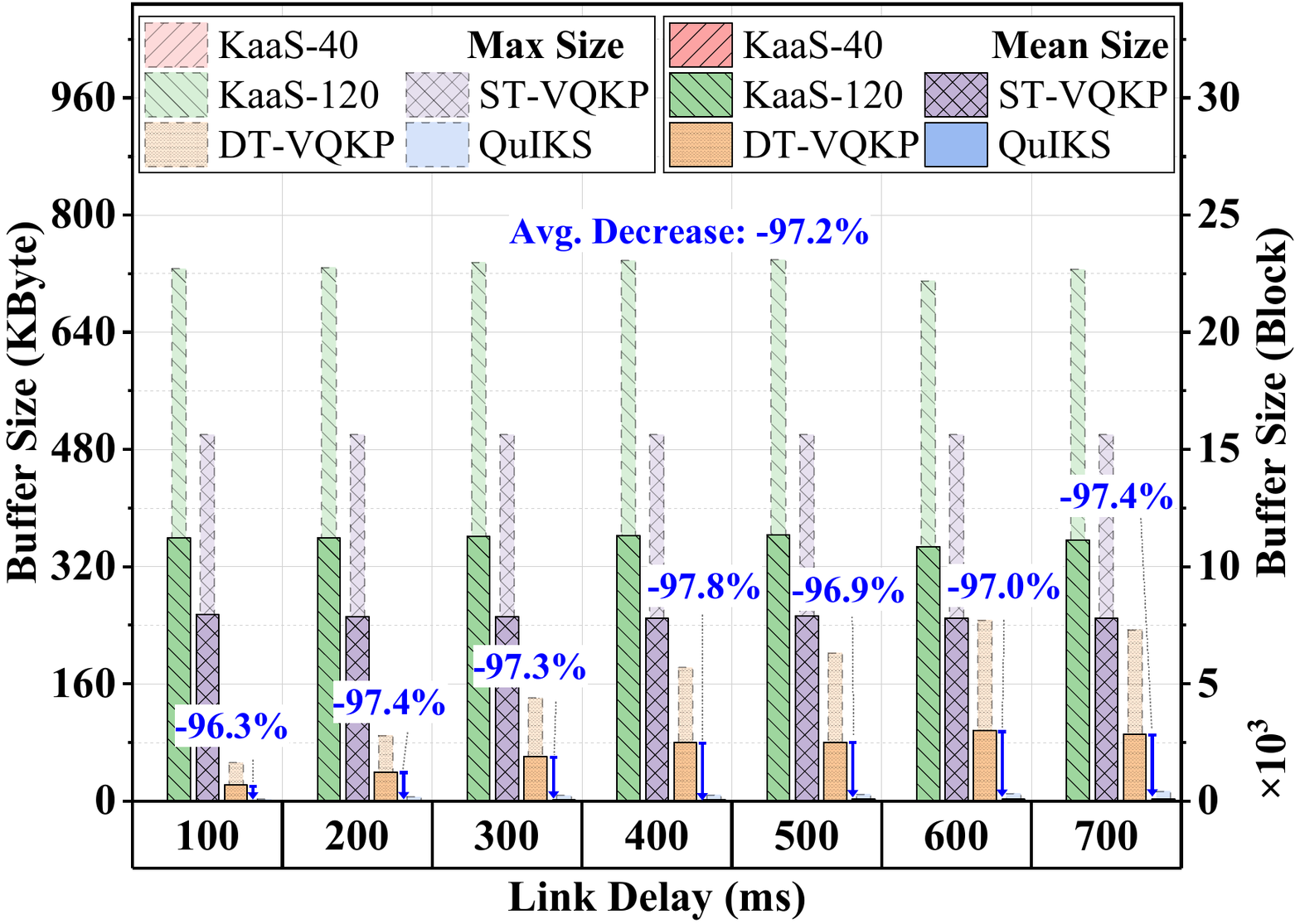}
    \subcaption{Max and mean buffer size.}\label{fig_bufpoi}
  \end{minipage}
  \caption{Performance when application requests arrive as a Poisson process in the single-application scenario.}\label{fig_poi}
\end{figure}

In this scenario, experiments are conducted between node 0 and node 1 with only one application to evaluate the basic performance of QuIKS and comparison schemes.
The application request rate is 50 rps, with the process being a Poisson process or a Poisson-Pareto Burst process (PPBP) \cite{akhtar2023fast}.
In the composite PPBP process, the sending event follows a Poisson process with an arrival rate of 1, and the number of sent requests follows a Pareto distribution with a shape parameter of 2.
The key demands of the application are 500 KByte, i.e., 15625 blocks, and quantum keys are sufficiently abundant in the network. The mean link delay ranges from 100 ms to 700 ms.

\cref{fig_poi} presents the experimental results when application requests arrive as a Poisson process.
As can be seen in \cref{fig_ksdpoi}, there are significant differences in key supply latency among NOBUFFER, KaaS-40, and other schemes.
For the NOBUFFER scheme, the key supply latency is directly affected by the link delay.
Due to a relaying rate lower than the application request rate, KaaS-40 always has an empty buffer as depicted in \cref{fig_bufpoi}, resulting in high and unstable key supply latency.
For other schemes like QuIKS, the key supply latency for 95\% of requests consistently remains lower than 0.5 ms, with some high outliers caused by a temporal empty buffer at startup, no matter how the link delay changes.
However, \cref{fig_bufpoi} shows that KaaS-120, ST-VQKP, and DT-VQKP have a large amount of buffer size, with the mean value up to 360 KByte achieved by KaaS-120.
The best one of them, DT-VQKP, has a lower mean buffer size down to 22 KByte, which grows with the increase of the link delay because its relaying strategy employs a product of the application request rate and link delay, multiplied by a large factor.
Even in this case, QuIKS's buffer size is still 97.1\% lower than that of DT-VQKP on average, demonstrating its ultra-low key consumption and high resource efficiency.

\begin{figure}[t]
  \centering
  \begin{minipage}[t]{0.48\linewidth}
    \centering
    \includegraphics[width=\linewidth]{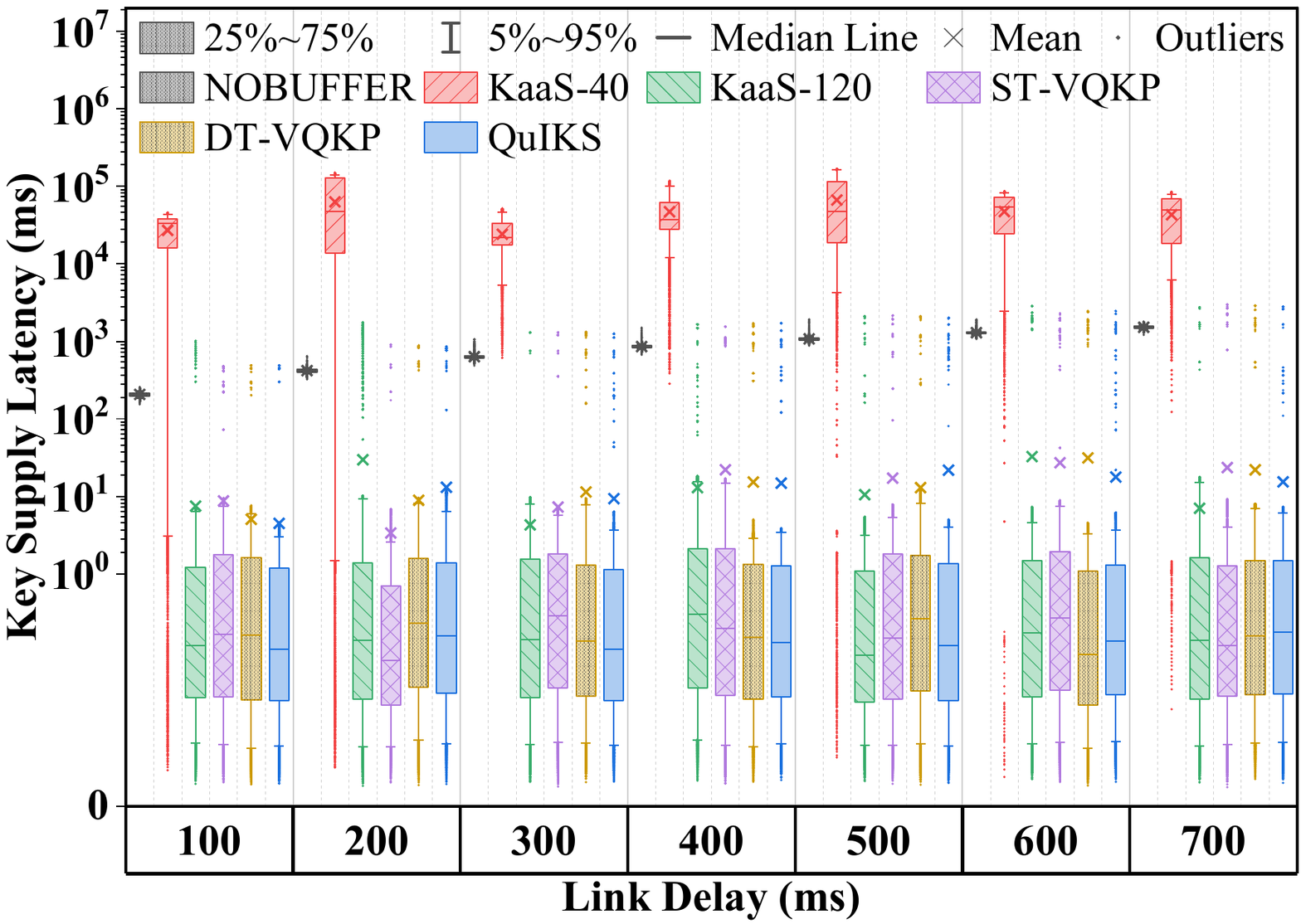}
    \subcaption{Key supply latency.}\label{fig_ksdppbp}
  \end{minipage}~
  \begin{minipage}[t]{0.48\linewidth}
    \centering
    \includegraphics[width=\linewidth]{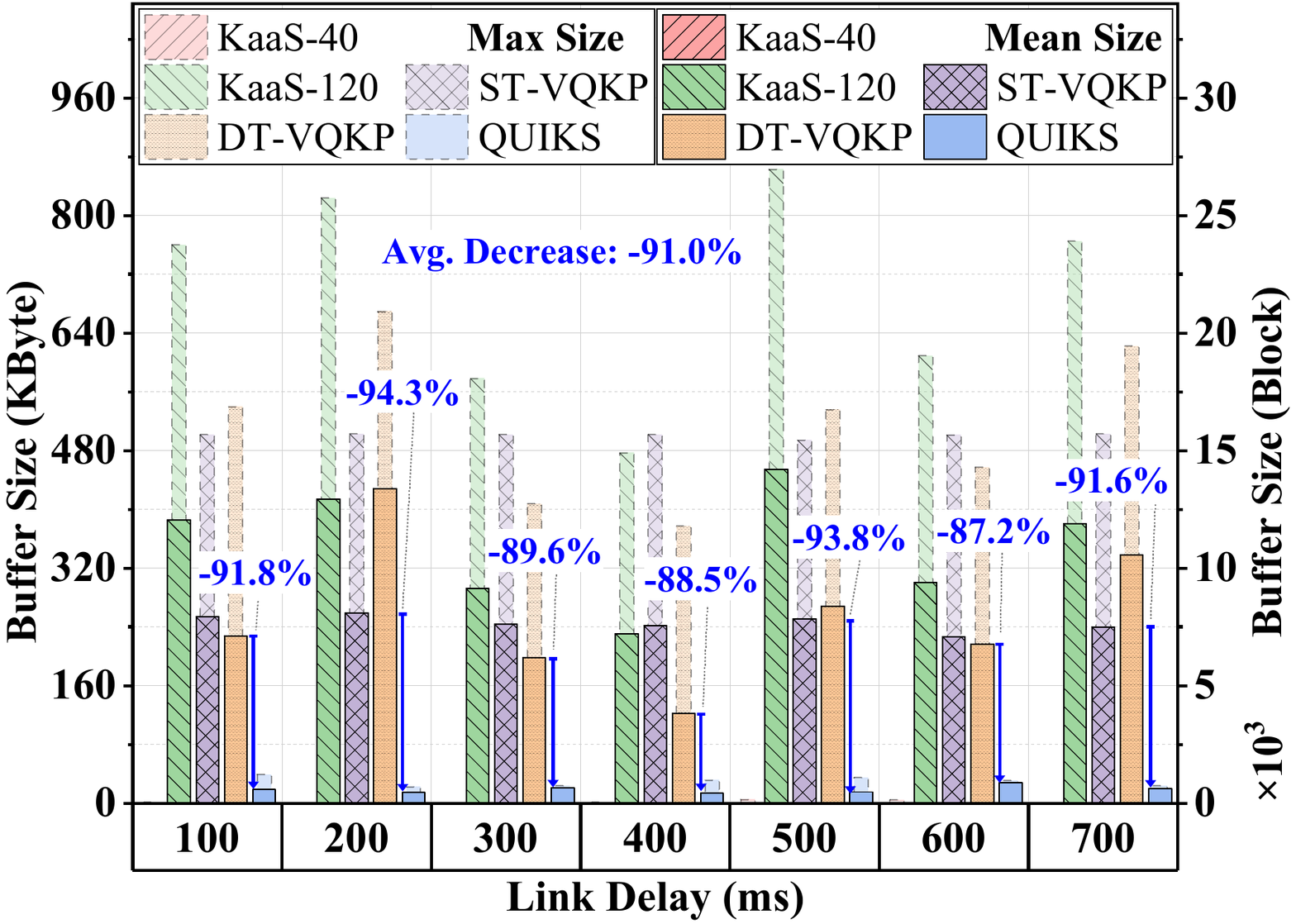}
    \subcaption{Max and mean buffer size.}\label{fig_bufppbp}
  \end{minipage}
  \caption{Performance when application requests arrive as a Poisson-Pareto process in the single-application scenario.}\label{fig_ppbp}
\end{figure}

\cref{fig_ppbp} illustrates the results when application requests arrive as a PPBP process, where application requests arrive in a bursty pattern.
In this case, the key supply latency increases and has a broader distribution as illustrated in \cref{fig_ksdppbp}.
Despite this, for schemes except NOBUFFER and KaaS-40, the key supply latency for 95\% of requests remains lower than 10 ms, where the increases are due to the bursty request pattern, independent of schemes.
For the buffer size in this scenario, \cref{fig_bufppbp} depicts a slightly different result.
In this scenario, among the schemes except QuIKS, the lowest mean buffer size is 122 KByte achieved by DT-VQKP when the link delay is 400 ms. ST-VQKP remains a larger but stable mean buffer size of around 250 KByte, as it is only related to the total number of requests, while KaaS-120 and DT-VQKP exhibit a random value of mean buffer size owing to the significant application request variations under the PPBP process.
Although QuIKS's buffer size increases compared with that under the Poisson process, it still remains over 90\% lower than that of the best one among other schemes on average, further demonstrating its advantage.

\subsection{Deeper Analysis of the Buffer Size}

\begin{figure}[t]
  \centering
  \begin{minipage}[t]{0.48\linewidth}
    \centering
    \includegraphics[width=\linewidth]{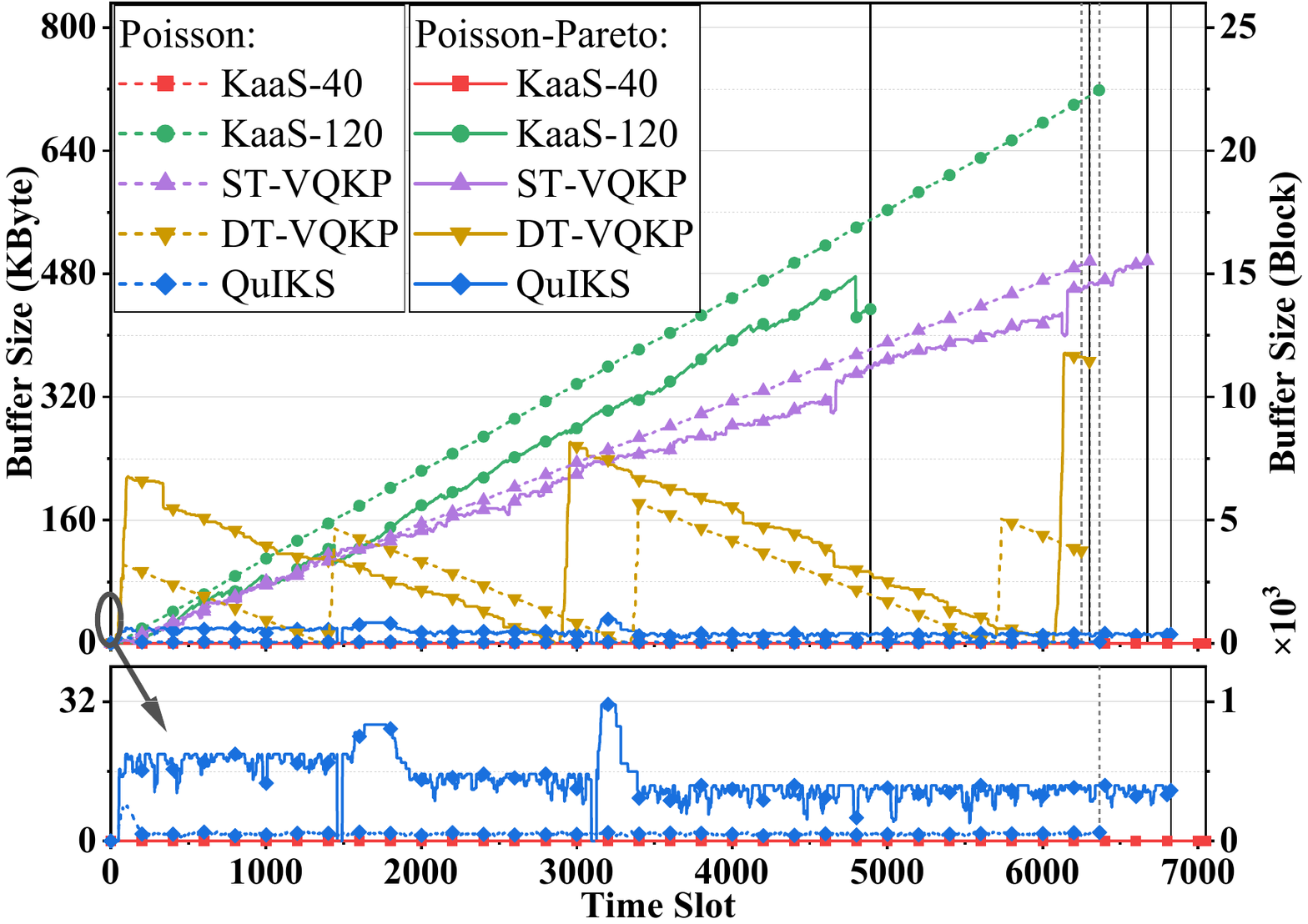}
    \caption{Buffer size per time slot when delay is 400 ms.}\label{fig_buf400}
  \end{minipage}~
  \begin{minipage}[t]{0.48\linewidth}
    \centering
    \includegraphics[width=\linewidth]{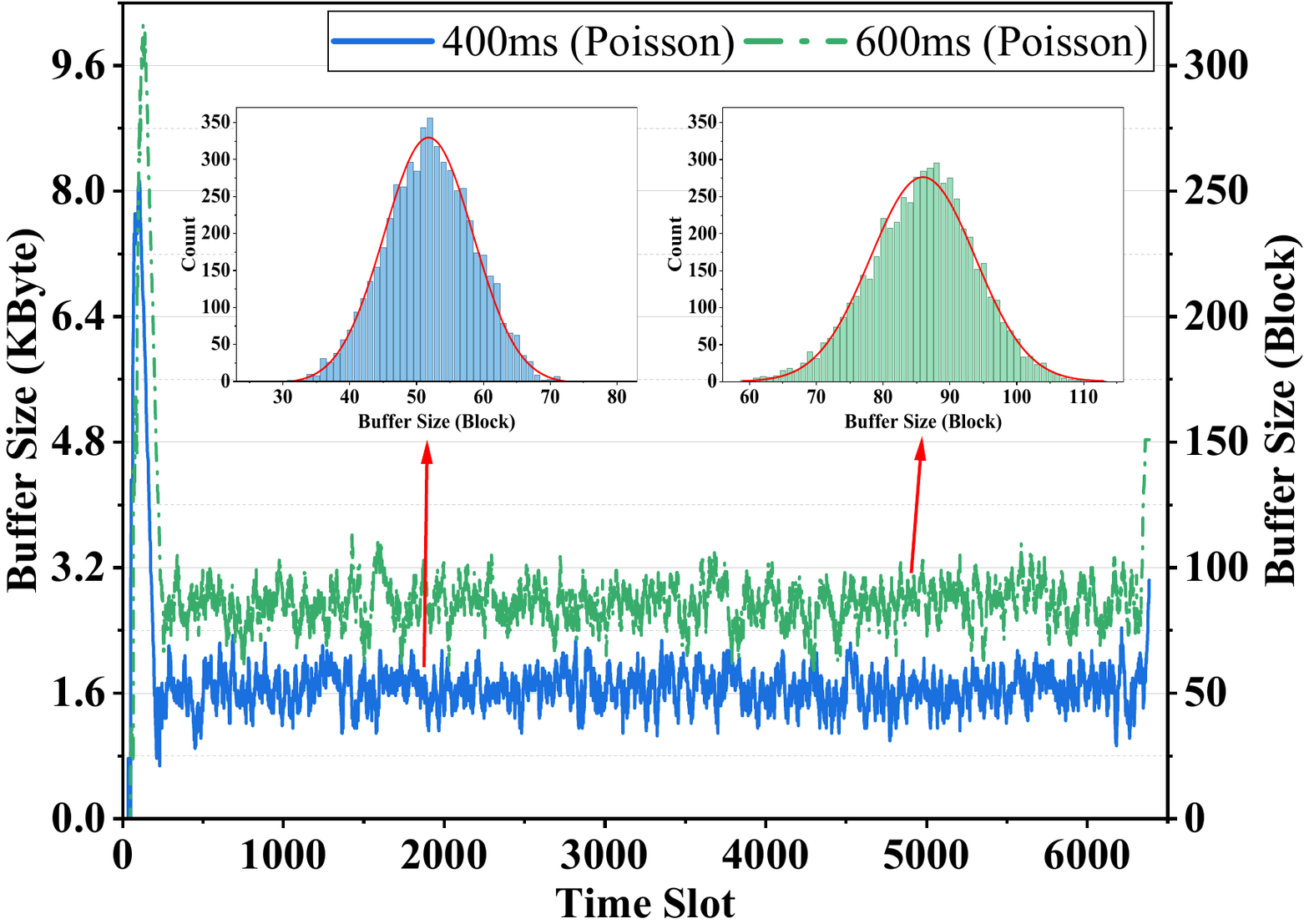}
    \caption{Fitting of the buffer size distribution for QuIKS.}\label{fig_buffit}
  \end{minipage}
\end{figure}

\cref{fig_buf400} uses data with a link delay of 400 ms as an example to demonstrate the key buffer size per time slot, showcasing the control mechanisms of each scheme.
These schemes, except KaaS-40 with zero buffer, can provide uninterrupted key supply.
Hence, their application completion times are also annotated to reflect the application request characteristics under either a Poisson or PPBP process, where these two types are represented by vertical gray dashed and black solid lines, respectively.
Under the PPBP process, the significant variation in application completion times is attributed to the high variation in request rate per time slot.
As shown in \cref{fig_buf400}, the buffer size of KaaS-120, with its constant relaying rate, is primarily affected by application completion time, while that of DT-VQKP, with a request rate-based bursty relaying method, is governed by the request rate at the moment of relaying.
Consequently, both exhibit larger and more fluctuating buffers as previously shown in \cref{fig_bufppbp}.
Furthermore, it can be seen that QuIKS's buffer size under the PPBP process is significantly larger than that under the Poisson process, with only a few occasional probing and adjusting phases during operation. This indicates that QuIKS successfully captures the drastic variations in bursty application requests and thus possesses high adaptiveness.

\begin{table}[t]
  \centering
  \caption{Comparison of parameters for QuIKS}\label{tab_fitpara2}
  \setlength{\aboverulesep}{1pt}
  \setlength{\belowrulesep}{1pt}
  \begin{tabular}{@{\hspace{3pt}}c@{\hspace{3pt}}|@{\hspace{2pt}}c@{\hspace{2pt}}c@{\hspace{1pt}}c@{\hspace{3pt}}c@{\hspace{6pt}}c@{\hspace{3pt}}}
    \toprule
    Delay & Real $\sigma_{\Delta}$ & Est. $\hat{\sigma}_{\Delta}$\textbf{($-\sigma_{\Delta}$)} & Fitted $\overline{\sigma}_{\Delta}$\textbf{($-\sigma_{\Delta}\%$)} & $R^2$ & $\chi^2_{red}$ \\
    \midrule
    100 ms & 3.4950 &  4.6596 \textbf{(+1.1646)} & 3.4342 \textbf{($-$1.7\%)} & 0.9887 & 2.4837 \\
    200 ms & 5.0575 &  5.9812 \textbf{(+0.9237)} & 4.8356 \textbf{($-$4.4\%)} & 0.9709 & 4.4418 \\
    300 ms & 5.6467 &  6.8012 \textbf{(+1.1545)} & 5.4672 \textbf{($-$3.2\%)} & 0.9785 & 2.6905 \\
    400 ms & 6.5996 &  7.5748 \textbf{(+0.9752)} & 6.6602 \textbf{($+$0.9\%)} & 0.9823 & 2.2301 \\
    500 ms & 7.6348 &  8.2602 \textbf{(+0.6254)} & 7.6662 \textbf{($+$0.4\%)} & 0.9904 & 1.0049 \\
    600 ms & 7.9527 &  9.0477 \textbf{(+1.0950)} & 7.8472 \textbf{($-$1.3\%)} & 0.9869 & 1.2471 \\
    700 ms & 8.9418 & 10.1152 \textbf{(+1.1734)} & 8.9471 \textbf{($+$0.1\%)} & 0.9846 & 1.3011 \\
    \bottomrule
  \end{tabular}
\end{table}

\cref{fig_buffit} illustrates the buffer size and the fitted distribution of QuIKS under a Poisson process, with link delays of 400 ms and 600 ms.
Discarding the initial several time slots of the probing and adjusting phase, QuIKS can maintain an ultra-low buffer size under both conditions, and its distribution intuitively fits a normal distribution.
This observation is validated in \cref{tab_fitpara2}, where the fitted $\overline{\sigma}_\Delta$ deviates from the real $\sigma_\Delta$ of the buffer size by less than 5\% and exhibits a high coefficient of determination ($R^2$) and low reduced chi-square $\chi^{2}_{red}$.
These results mean that the distribution of buffer size statistically follows a normal distribution, which confirms the accuracy of \cref{remark_buffer}.
On this basis, the $\hat{\sigma}_\Delta$ estimated according to \cref{alg_probe} exhibits a stable positive difference of around $1$ to the real $\sigma_\Delta$, meaning that it can capture the most important trends in buffer size variation when link delay changes, despite the bias due to lightweight designs. Also, the overestimation provides a conservative margin in practice, while small enough to keep the superiority of QuIKS over other schemes.

\subsection{Performance in a Multiple-Application Scenario}

\begin{figure}[t]
  \centering
  \begin{minipage}[t]{0.48\linewidth}
    \centering
    \includegraphics[width=\linewidth]{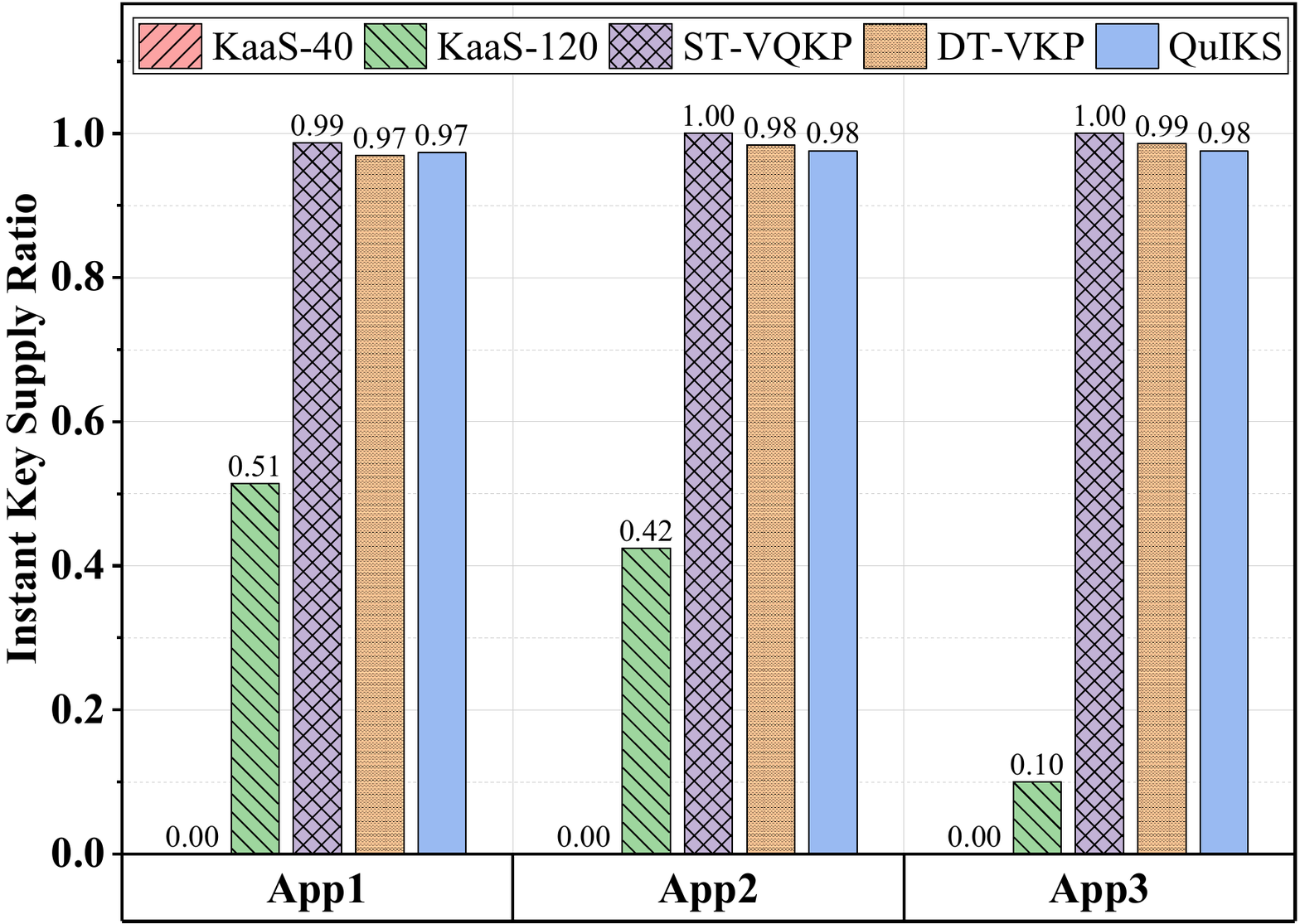}
    \subcaption{Instant key supply ratio.}\label{fig_appiksr}
  \end{minipage}~
  \begin{minipage}[t]{0.48\linewidth}
    \centering
    \includegraphics[width=\linewidth]{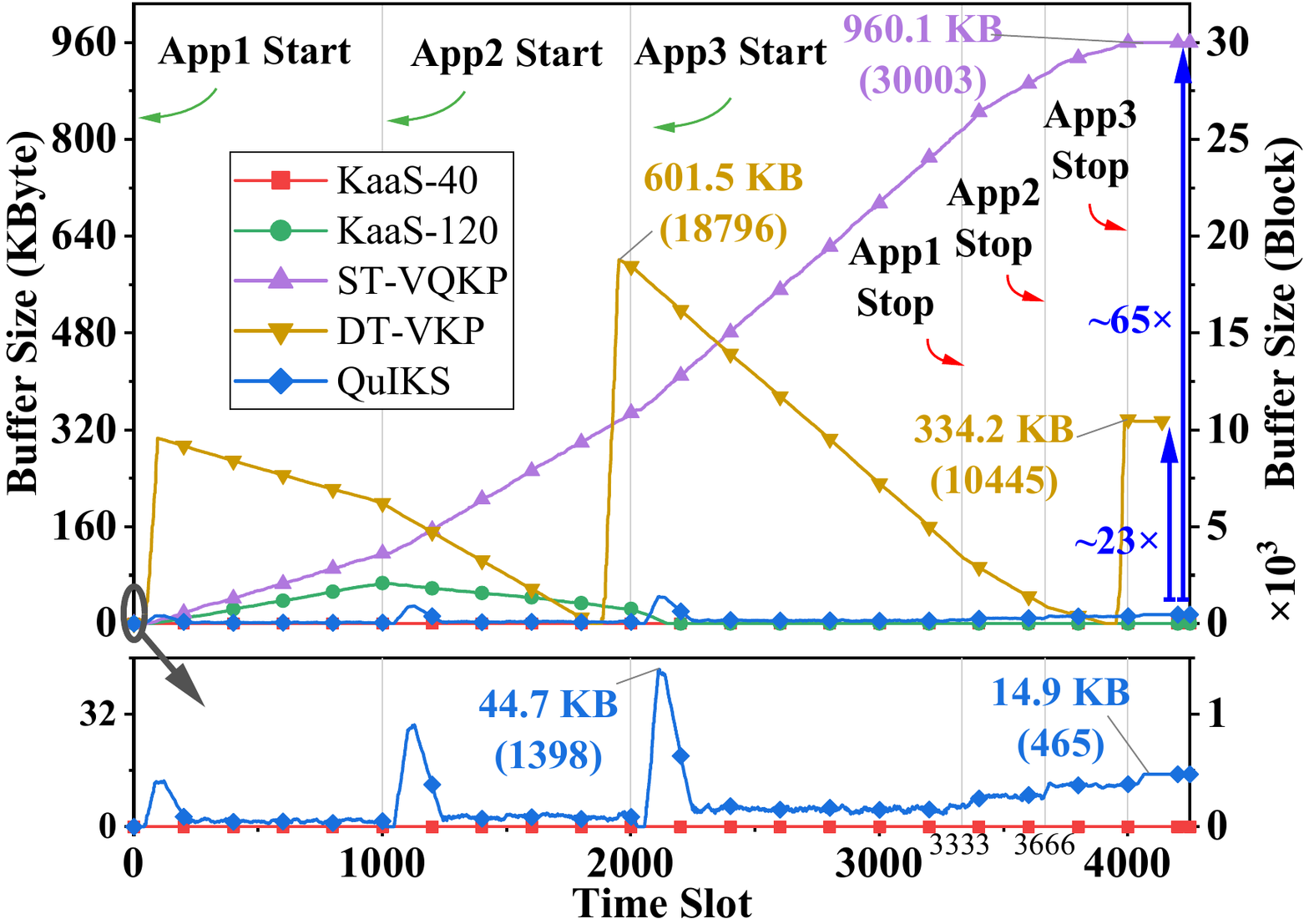}
    \subcaption{Buffer size per time slot.}\label{fig_appbuf}
  \end{minipage}
  \caption{Performance in the multiple-application scenario.}\label{fig_app}
\end{figure}

In this scenario, experiments are conducted between node 3 and node 7, following the path of 3-4-6-7, to evaluate how multiple applications affect the buffer. Three applications start in sequence, each with a request rate of 75 rps in a Poisson process and key demands of 400, 320, and 240 KByte, respectively. The mean delay for each link is set to 200 ms, and quantum keys are sufficiently abundant.

\cref{fig_app} presents the experiment results, where KaaS exhibits a poor instant key supply ratio in \cref{fig_appiksr}. This is attributed to its fixed relaying rate, which cannot cope with dynamic request rates.
As shown in \cref{fig_appbuf}, when there is only one application, KaaS-120 can handle and accumulate some keys in the buffer.
But as the subsequent applications join, the total request rate (\textgreater150) exceeds its relaying rate, making the buffer eventually be exhausted and resulting in a significant drop in performance after time slot 2000.
For ST-VQKP, its relaying rate is always twice the request rate, so there is a clear change in buffer size growth when the application starts or stops. Hence, it leads to an optimal instant key supply ratio but an extremely large buffer size.
However, as illustrated in \cref{fig_appiksr}, both DT-VQKP and QuIKS achieve a comparable ratio of over 0.97 while reducing the buffer size.
Notably, QuIKS reduces buffer size at the end by $65\times$ and $23\times$ over ST-VQKP and DT-VQKP, respectively, with a temporally large size occurring only when applications join.
Therefore, QuIKS can adapt to multiple applications and still maintain an instant key supply with ultra-low key resource consumption.

\subsection{Performance in a Key-Limited Scenario}

In this scenario, to evaluate the performance of various schemes when faced with numerous applications and limited quantum keys, experiments are conducted across the entire NSFnet topology as depicted in \cref{fig_impl} with the shortest path routing.
Applications are generated with random positions, their quantity ranging from 20 to 80.
Each of them starts randomly within [0, 150] seconds, and has a request rate of 10 rps in a Poisson process and key demands of 50 KByte. The mean delay for each link is set to 200 ms, and quantum keys are merely 10\% higher than the minimum requirements.

\begin{figure}[t]
  \centering
  \begin{minipage}[t]{0.48\linewidth}
    \centering
    \includegraphics[width=\linewidth]{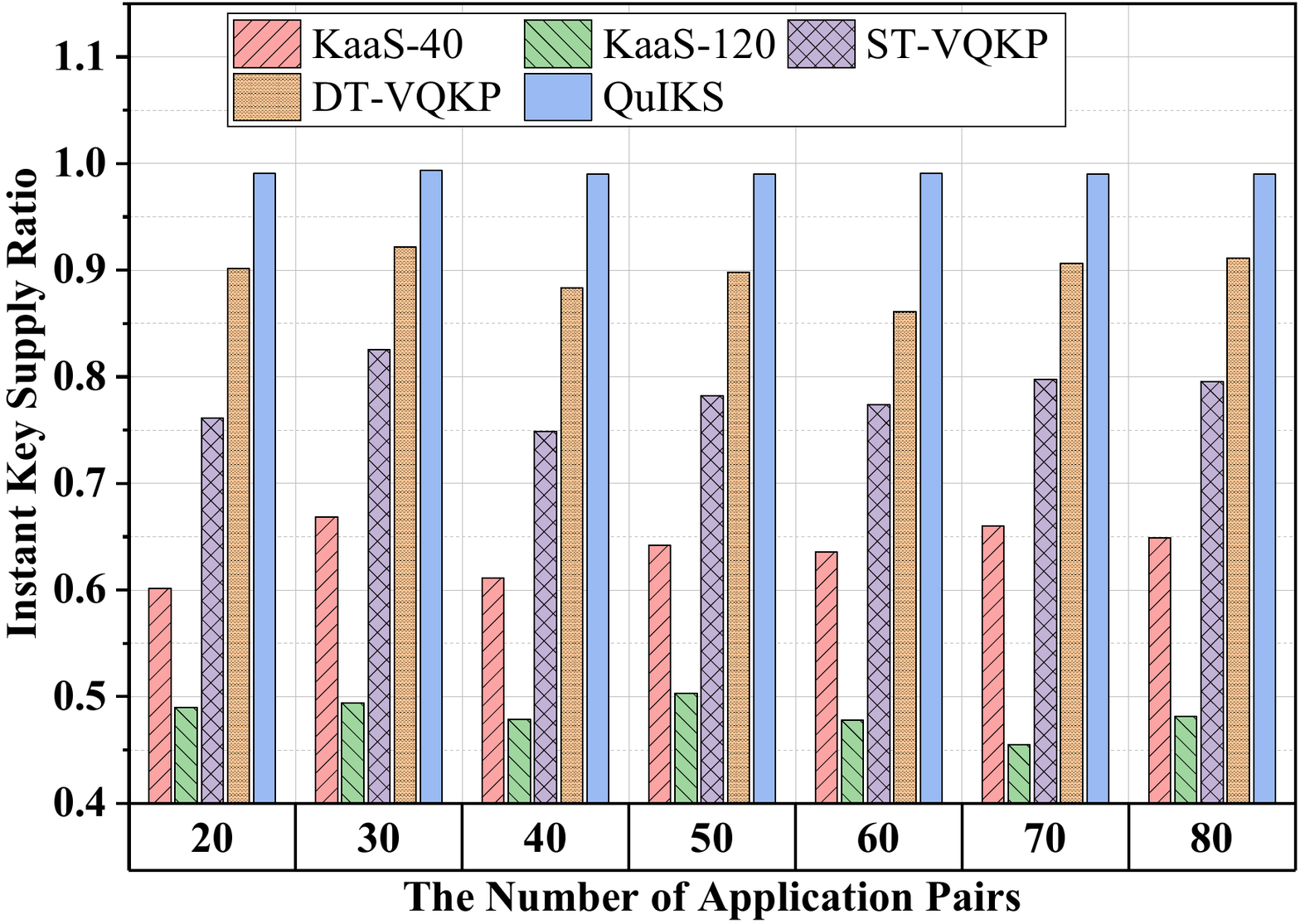}
    \subcaption{Instant key supply ratio.}\label{fig_nsfqos}
  \end{minipage}~
  \begin{minipage}[t]{0.48\linewidth}
    \centering
    \includegraphics[width=\linewidth]{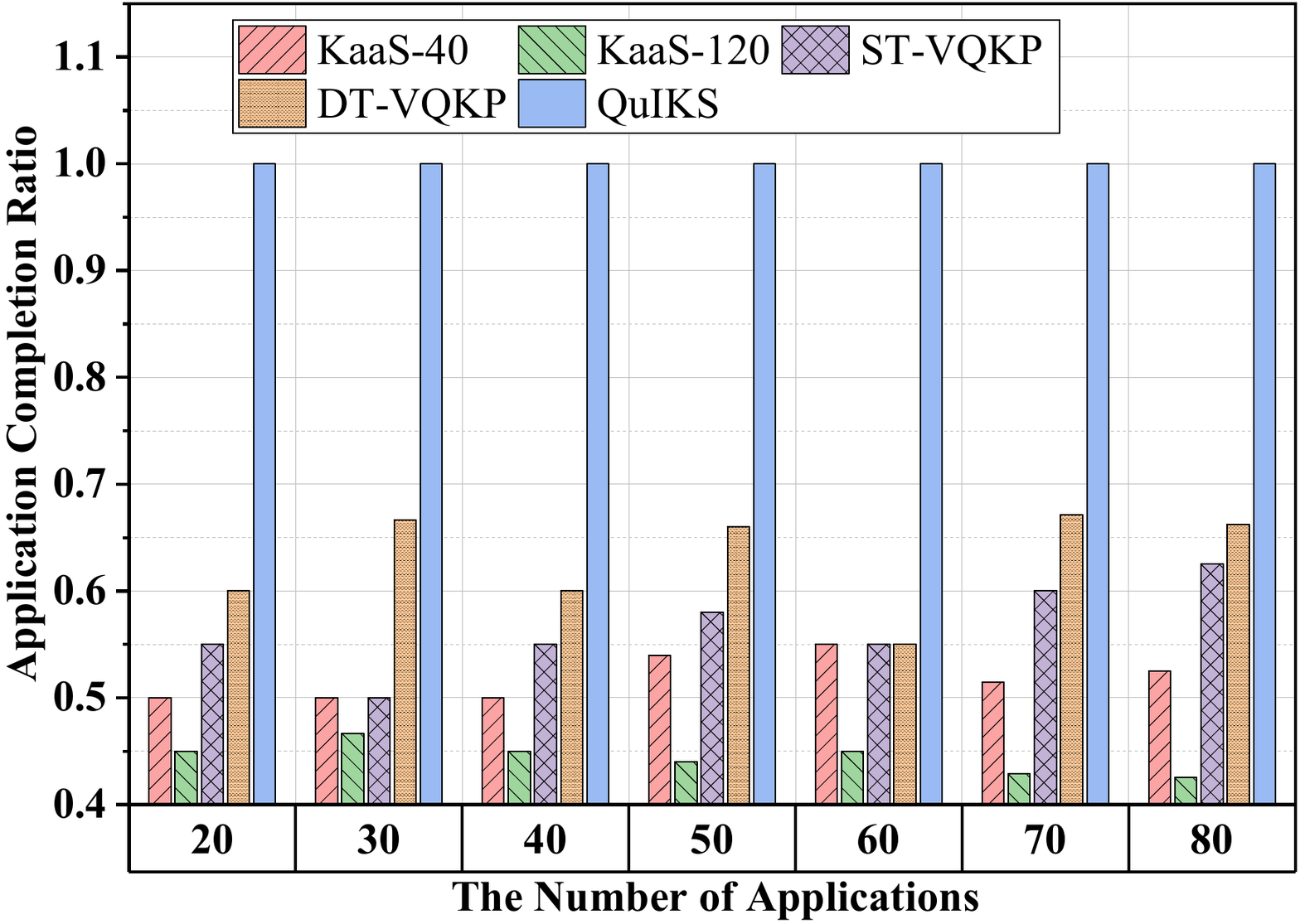}
    \subcaption{Application completion ratio.}\label{fig_nsfapp}
  \end{minipage}
  \caption{Performance in the key-limited scenario.}\label{fig_nsf}
\end{figure}

\cref{fig_nsfqos} illustrates the instant key supply ratio of all schemes.
In this case, KaaS-40 is better than KaaS-120 because KaaS does not stop relaying, and the latter has a higher relaying rate, which can easily exhaust the quantum keys in early-starting buffers.
Without available quantum keys, the key relaying necessary for key supply fails, resulting in a low application completion ratio, as depicted in \cref{fig_nsfapp}.
In this key-limited scenario, lower quantum key consumption actually leads to optimal performance.
ST-VQKP sends relaying requests at double the rate of application requests and stops with them.
Therefore, it consumes fewer quantum keys than KaaS, ensuring that late-starting buffers have sufficient quantum keys available, which leads to a higher instant key supply ratio and application completion ratio.
Similar to the results in previous scenarios, DT-VQKP exhibits sawtooth-like buffer fluctuations that do not continuously increase, thereby consuming much fewer quantum keys and achieving 90\% instant key supply ratio as shown in \cref{fig_nsfqos}. However, DT-VQKP still requires a large buffer size, leading to key supply interruptions for many applications nearing completion, as shown in \cref{fig_nsfapp}.
QuIKS achieves a 100\% application completion ratio and a near-optimal 99\% instant key supply ratio with an ultra-low buffer size and quantum key consumption. The remaining 1\% is attributed to the inherent queuing that occurs when a buffer is initializing. The results further demonstrate that QuIKS offers an excellent trade-off between performance and key resource consumption, meeting the design goals.

\section{Related Work}\label{works}

\textbf{Path-level key management.}
A primary field of research focuses on maximizing key utilization efficiency on a single path to provide sufficient key resources for continuous instant key supply.
Protocol-level innovations, such as key caching in IC-QKD \cite{zhang2023icqkd} and precise key and request management in AKRP \cite{chen2025asynchronous}, aim to reduce key consumption during the relaying process.
From a resource allocation perspective, some works \cite{zhou2022quantum,li2025decentralized} introduce management schemes that prioritize critical requests to enhance overall relaying performance on each node.
However, these path-level optimizations are fundamentally constrained by the finite key resources available on that single path.
They cannot aggregate key resources in the network, making them insufficient when a path's intrinsic quantum keys cannot meet escalating demands.

\textbf{Network-wide path selection.}
To overcome the limitations of single-path schemes, numerous studies have explored dynamic path selection to aggregate key resources across the QKD network.
These routing schemes dynamically discover key-abundant paths to serve requests \cite{mehic2019novel,xu2024routing,zheng2025integration,wenning2025comparison}.
Some advanced approaches also incorporate classical network metrics, such as latency and bandwidth, to bypass underperforming paths \cite{mehic2019novel,akhtar2023fast}.
However, while solving for quantum key availability, these schemes introduce new performance volatility. The switching between paths with heterogeneous characteristics (e.g., latency, hops), coupled with the unpredictable performance of classical networks, where key relaying is executed, leads to an unstable key distribution.
Consequently, they fail to guarantee the stable and predictable relaying performance required for instant key supply.

\textbf{Service-layer decoupling via buffering.}
To fundamentally address the problem, end-to-end buffering has emerged as a promising paradigm to decouple key supply services from network volatility.
The concept is pioneered in KaaS \cite{cao2019kaas}, which uses static scheduling to pre-fill buffers.
Subsequent works have refined this idea, from immediately relaying keys to construct application-specific buffers \cite{yu2021multi}, to hierarchical slicing for finer-grained services \cite{zhu2023qkd}.
More recently, research has begun to explore dynamic buffering strategies \cite{stan2025dynamic}.
However, while establishing the benefits of buffering, these existing schemes predominantly rely on static or heuristic methods for determining buffer sizes and relaying strategies.
A systematic framework that balances the trade-offs between key supply performance and overhead, such as key consumption, remains a critical and under-addressed research gap.

\section{Conclusion}\label{conclu}

In this paper, we proposed an instant key supply scheme based on adaptive buffering in QKD networks, named QuIKS, to address the dilemma between key resource consumption and key supply performance.
The challenges stem from the fundamental difficulty of modeling stochastic buffer dynamics and the lack of a quantitative framework.
To overcome these challenges, we established a novel theoretical model to analyze the impact of stochastic application requests and the QKD network on buffer dynamics, and then we quantified the mathematical relationship between key supply performance and buffer size.
Guided by this model, we designed a lightweight two-phase buffer control algorithm to provide instant key supply services with a theory-guided ultra-low key buffer size.
Extensive experiments on a real-world QKD testbed demonstrate that QuIKS achieves a near-zero key supply latency while reducing key buffer size by more than 90\%, thus minimizing key resource consumption and paving the way for high-performance and resource-efficient QKD networks.

\section*{Acknowledgement}
This work is supported in part by the National Natural Science Foundation of China under Grant No. 62572450, Grant No. 62402466, and Grant No. 62501562,
the Innovation Program for Quantum Science and Technology under Grant No. 2021ZD0301301,
the Youth Innovation Promotion Association Chinese Academy of Sciences under Grant No. Y202093,
and Japan Society for the Promotion of Science (JSPS) KAKENHI under Grant No. 23K28070.

\flushcolsend

\clearpage

\bibliographystyle{IEEEtran}
\bibliography{references}

\end{document}